# First supra-THz Heterodyne Array Receivers for Astronomy with the SOFIA Observatory

Christophe Risacher, Rolf Güsten, Jürgen Stutzki, Heinz-Wilhelm Hübers, Denis Büchel, Urs U. Graf, Stefan Heyminck, Cornelia E. Honingh, Karl Jacobs, Bernd Klein, Thomas Klein, Christian Leinz, Patrick Pütz, Nicolas Reyes, Oliver Ricken, Hans-Joachim Wunsch, Paul Fusco, Stefan Rosner

*Abstract*— We present the upGREAT THz heterodyne arrays for far-infrared astronomy. The Low Frequency Array (LFA) is designed to cover the 1.9-2.5 THz range using 2x7-pixel waveguide-based HEB mixer arrays in a dual polarization configuration. The High Frequency Array (HFA) will perform observations of the [OI] line at ~4.745 THz using a 7-pixel waveguide-based HEB mixer array. This paper describes the common design for both arrays, cooled to 4.5 K using closed-cycle pulse tube technology. We then show the laboratory and telescope characterization of the first array with its 14 pixels (LFA), which culminated in the successful commissioning in May 2015 aboard the SOFIA airborne observatory observing the [CII] fine structure transition at 1.905 THz. This is the first successful demonstration of astronomical observations with a heterodyne focal plane array above 1 THz and is also the first time high-power closed-cycle coolers for temperatures below 4.5 K are operated on an airborne platform.

## I. INTRODUCTION:

WITH the successful Herschel observatory satellite mission, which was active between 2009-2012 [1], astronomical observations over the 1-5 THz range were performed with order of magnitude of improved sensitivities, compared to previous satellite missions. This far-infrared region cannot be explored from ground-based telescopes (with the exception of narrow frequency windows observable from high altitude locations). The instrumentation for this wavelength range is divided into two types, direct detection cameras which can have low to mid-resolution spectroscopic capabilities, and heterodyne receivers which provide very high resolution spectroscopy. The main reason for wanting high resolution is to resolve the spectral lines profiles, which allows studying the gas excitation and kinematics in great detail.

The continuum-type instrumentation routinely incorporates several hundred or more pixels and there is rapid progress in this area [2-4]. For high resolution spectroscopy using heterodyne receivers, only recently did the detector sensitivities start approaching their fundamental noise limit up to very high frequencies [5-6] and the next step is now to increase the receiver pixels count in order to gain observing efficiency when observing extended sources. The review papers [7-8] summarize the current status of multi-pixel heterodyne array receivers for the sub-mm, far-infrared region and present general design considerations. Until now, only small sized arrays have been realized in a handful of observatories for frequencies between 0.3-0.8 THz (e.g. [9-12]) and only single pixel receivers have been successfully operated on telescopes for frequencies above 1 THz (e.g. [13-15]). Several THz array projects are ongoing and concepts have been presented for THz heterodyne arrays (e.g. [16-18]) but until now, none has been successfully operated on a telescope. Several satellite missions for the far infrared region (SPICA [19], Millimetron [20]) and balloon experiments (STO-2 as successor of STO [13] or GUSTO) plan to use array instruments in the future. Currently the SOFIA[1] airborne observatory, a NASA/DLR operated Boeing 747 carrying a 2.5 m telescope, is the only platform capable of observing in that wavelength region [21].

Among the SOFIA instruments, the only one able to

Submitted on August 31st 2015. This work was supported in part by the Federal Ministry of Economics and Technology via the German Space Agency (DLR) under Grants 50 OK 1102, 50 OK 1103 and 50 OK 1104 and by the Collaborative Research Council 956, sub-projects D3 and S, funded by the Deutsche Forschungsgemeinschaft (DFG).

C. Risacher, R. Güsten, S. Heyminck, B. Klein, T. Klein, O. Ricken are with the Max Planck Institut für Radioastronomie, Bonn, 53121 Germany (e-mail: crisache@mpifr.de).
B. Klein is also with the University of Applied Sciences Bonn-Rhein-Sieg, Sankt Augustin, 53757 Germany (bklein@mpifr.de).
T. Klein is also with the European Southern Observatories, Vitacura, Santiago de Chile, 19001, Chile (e-mail: tklein@eso.org).
N. Reyes was with the Max Planck Institut für Radioastronomie, Bonn, 53121 Germany and is now with the Universidad de Chile (e-mail: nireyes@u.uchile.cl).
J. Stutzki, P. Pütz, C. Honingh, U. U. Graf, K. Jacobs, D. Büchel are with the Cologne University, 50937, Köln, Germany (e-mail: stutzki@ph1.uni-koeln.de).
H.-W. Hübers is with the German Aerospace Center (DLR), Institute of Optical Sensor Systems, 12489 Berlin, Germany (email: Heinz-Wilhelm.Huebers@dlr.de).
P. Fusco is with NASA Ames Research Center, Moffett Field, CA 94035, USA (paul.r.fusco@nasa.gov).
S. Rosner is with the SETI Institute, Mountain View, CA 94043, USA (e-mail: stefan.rosner@nasa.gov).

---

[1] SOFIA is a joint project of NASA and the German Aerospace Center (DLR). The aircraft is based at the NASA Armstrong Flight Research Center (AFRC) facility in Palmdale, California which also manages the program. NASA Ames Research Center at Moffett Field, California, manages the SOFIA science and mission operations in cooperation with the Universities Space Research Association (USRA) headquartered in Columbia, Maryland, and the German SOFIA Institute (DSI) at the University of Stuttgart.

achieve high resolution spectroscopy (above $10^6$ resolving power) is the GREAT instrument [22]. In this paper, we present the new array receivers for this instrument, the upGREAT multi-pixel heterodyne arrays for frequencies 1.9-2.5 THz, and 4.75 THz.

## II. THE GREAT/UPGREAT RECEIVERS

The GREAT instrument[1] is described in detail in [22]. Its modular construction allows using at any time two cryostats mounted on the main support structure, which is then mounted to the telescope Nasmyth flange (Fig.1). Both cryostat channels observe in parallel the same sky position.

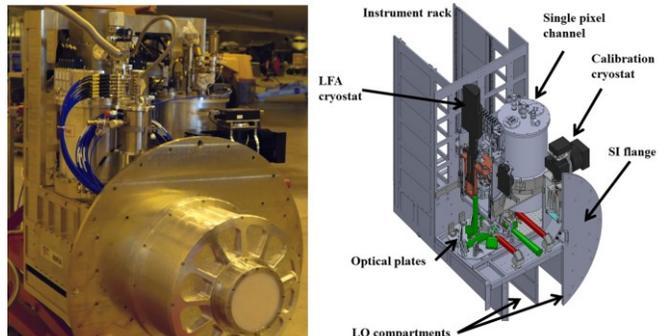

Fig. 1. In this example the GREAT instrument accommodates the upGREAT LFA channel (left cryostat) and a single pixel cryostat (L1 band) operated simultaneously. A smaller cryostat contains the 80K load used for calibration. The SI flange attaches to the telescope Nasmyth tube. The sky signal (in green) is separated either in polarization by a wire grid or in frequency by a dichroic filter. The local oscillator components are placed under the structure. The LO signals (in red) are coupled with the sky signal on an optical plate located inside the structure. Sensitive electronic components are located in the electronics rack section

TABLE I
GREAT RECEIVER PERFORMANCE AS OF 2011 AND 2015

| Bands | Characteristics | Performance May 2011 | Performance January 2015 |
|---|---|---|---|
| L1 | Frequency range | 1.25-1.5 THz | 1.25-1.50 THz |
|  | IF Bandwidth | 1.0-2.5 GHz | 0.2-2.5 GHz |
|  | Trec (DSB) | 1000 K-1750 K | 500 K@0.5GHz |
|  | LO coupling | Diplexer optics | Beam splitter optics |
| L2 | Frequency range | 1.81-1.91 THz | 1.81-1.91 THz |
|  | IF Bandwidth | 1.0-2.5 GHz | 0.2-2.5 GHz |
|  | Trec (DSB) | 1500 – 4000 K | 600 K@0.5GHz |
|  | LO coupling | Diplexer optics | Beam splitter optics |
| Ma | Frequency range | 2.49-2.52 THz | 2.49-2.52 THz |
|  | IF Bandwidth | 1.0-2.5 GHz | 1.0-2.5 GHz |
|  | Trec (DSB) | 3000 K | 1500 K@2GHz |
|  | LO coupling | Diplexer optics | Diplexer optics |
| H | Frequency range |  | 4.745 ± 0.004 THz |
|  | IF Bandwidth |  | 0.2-2.5 GHz |
|  | Trec (DSB) |  | 800 K@0.5GHz |
|  | LO Technology |  | QCL |
|  | LO coupling |  | Beam splitter optics |

The main advantage of operating an airborne observatory as opposed to a satellite is the flexibility in placing newer developed technologies with very short turn-around times. Table I lists the available receiver bands and performance when GREAT was first installed in 2011 and after several years of operation, in January 2015. These are all single pixel cryostats cooled with liquid nitrogen and liquid helium. A significant improvement can be seen in the receiver performance in Table I and is explained by various upgrades: 1) improved optics, 2) better hot electron bolometer (HEB) detectors 3) higher output power local oscillator solid state chains. Most noticeably, a new receiver band based on waveguide HEB mixers, was added in 2014 [23], opening the [OI] 4.7 THz band for high-resolution spectroscopy for the first time since the pioneering Kuiper Airborne Observatory (KAO) in 1988-1995 [24].

The new upGREAT receivers' main characteristics are listed in Table II. The Low Frequency Array (LFA) covers the 1.9-2.5 THz range using 2x7-pixel array (dual polarization). The High Frequency Array (HFA) observes at 4.745 THz with a single polarization 7-pixel array. Each receiver channel is integrated in a single closed-cycle cryostat.

TABLE II
UPGREAT SPECIFICATIONS

|  | LFA array | HFA array |
|---|---|---|
| RF Bandwidth | 1.9-2.5 THz | 4.745 ± 0.004 THz |
| IF Bandwidth | 0.2-4 GHz | 0.2-4 GHz |
| HEB technology | Waveguide feedhorn antenna coupling, NbN HEB on Si membrane | Waveguide feedhorn antenna coupling, NbN HEB on Si membrane |
| LO technology | Photonic mixers / solid-state chains | Quantum cascade lasers (QCL) |
| LO coupling | Beamsplitter (goal) or Diplexer (baseline) | Beamsplitter |
| Array layout | 2x7 pixels dual polar. in hexagonal layout with a central pixel | 7 pixels single polar. in hexagonal layout with a central pixel |
| $T_{REC}$ | Goal <1000K DSB | Goal <1500K DSB |
| Backends | 0-4 GHz with min. 32k channels | 0-4GHz with min. 32k channels |

This paper presents the main design, testing and verification for common parts of the LFA/HFA receivers and presents the LFA receiver final integration, laboratory characterization and subsequent commissioning campaign results. The cryocooler infrastructure aboard the Boeing 747 is also presented here.

## III. UPGREAT INSTRUMENT DESIGN

### A. General Description

The main components of upGREAT are illustrated in Fig. 1-2. As for any heterodyne receiver system, a high-precision reference signal called local oscillator (LO) is required. The LO and the sky signal from the telescope are combined on an optics plate and are coupled to the superconducting HEB mixers. The intermediate frequency (IF) signal is amplified by cryogenic microwave low noise amplifiers (LNA) placed directly at the mixers output. The mixers and LNAs are located inside the cryostat and cooled to temperatures below 4.5 K. The IF signals are brought to the cryostat output connectors by stainless steel coaxial lines and are further amplified by room temperature amplifiers. The IF processor provides additional gain, equalization, filtering and computer-controlled output leveling of the IF signal, to match the nominal input level of the digital spectrometers.

In our case, the most demanding configuration is when the

---
[1] GREAT is a development by the MPI für Radioastronomie (Principal Investigator: R. Güsten) and the KOSMA/Universität zu Köln, in cooperation with the MPI für Sonnensystemforschung and the DLR Institut

LFA and HFA arrays are used in parallel. In that case, up to 21 HEB mixers, cryogenic LNAs, warm amplifiers and IF signal chains are active (shown in Fig. 2). Until May 2015, at most two channels were used on the GREAT instrument, therefore all of the bias electronics, IF processor, spectrometers were redesigned and upgraded to accommodate a maximum of 21 channels.

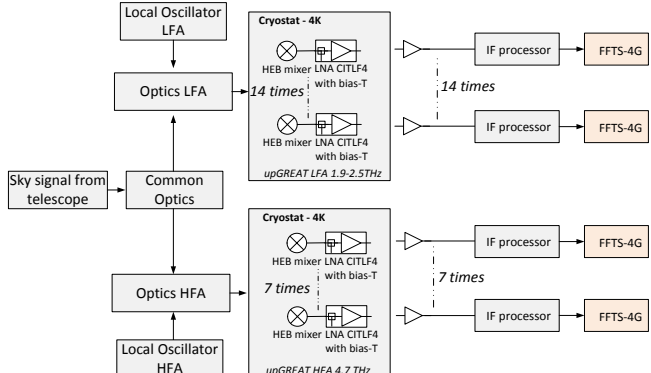

Fig. 2. General layout of the LFA/HFA configuration, when 14+7 signal chains are then needed.

### B. upGREAT LFA Local Oscillator

#### 1) Introduction

The upGREAT LFA has a broad RF range going from 1.9 to 2.5 THz and all of the components are designed to cover that range. The limiting factor at the moment is the availability of suitable LO sources with sufficient output power and RF bandwidth.

Solid-state multiplying chains are the current sources of choice, providing several tens of µW at 1.9 THz and with relative bandwidths of ~10%. The other candidate as a source is the Quantum Cascade Laser (QCL) which can produce several hundreds of µW power [25-26]. In parallel, we are developing in-house at the MPIfR photonic mixer local oscillators to cover the full RF range, from 1.9 to 2.5 THz. Photomixer-based sources have demonstrated several µW of power at 1 THz [27]. Despite this promising result, this is not sufficient yet to pump our HEB detectors with a straightforward and broadband beam-splitter coupling optics. Once the photonic LOs demonstrate sufficient LO power (a minimum of a factor of 2 increase is needed), they will be used with the LFA receiver in subsequent flight series (one photo-mixer per HEB mixer). The modularity of the system allows an easy replacement of the existing LO module with future improved units.

#### 2) General description

For the first observing campaigns with the SOFIA observatory, the LO sources are solid state multiplying chains, developed by Virginia Diodes, Inc. [28]. Their main characteristics are presented in Table III. The frequency coverage is limited to 1.88 to 1.92 THz, centered to cover the [CII] transition. Those LO chains use novel technologies, such as diamond substrates for thermal management, in-phase combining networks, and power amplification at 30 GHz to achieve an output power of 20 µW at ambient temperature. To further increase the output power, the last two passive triplers are cooled to 80 K, effectively doubling the output power to about 40 µW.

TABLE III
upGREAT LFA LO Optical parameters

| | |
|---|---|
| LO Bandwidth | 1.882-1.920 THz (limited by the LO output power) |
| Output power | 20 µW at ambient temperature |
| | 40 µW when last triplers are cooled to 80K |
| Optical coupling | Fourier Phase grating separates output beam into 7 equal amplitude beams |

In order to provide sufficient LO power for both LFA sub-arrays (2x7 HEB mixers), two identical LO chains are used, one per polarization. The outputs for both chains have orthogonal polarizations and are combined via a wire grid located inside the LO cooler (Fig. 3, 5). The resulting output LO beam is then split into 7 equal beams by using a Fourier phase grating [29] designed to operate at a center frequency of 1.9 THz with about 10% usable bandwidth. Every output beam contains about 12.8% of the incident power (90% efficiency).

All of the LO adjustment optics and phase grating are therefore common to the combined LO signals. They are separated afterwards by another wire grid, located on the optical plate and couple afterwards to their respective LFA sub-arrays.

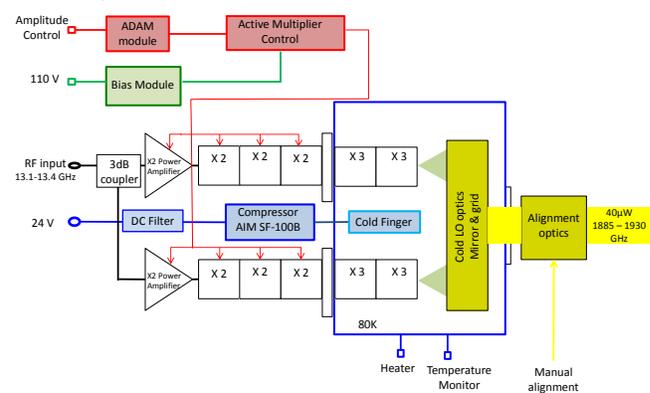

Fig. 3. LFA Local Oscillator general diagram. The identical chains, AMC-410 and AMC-461 produce about 20 µW power at 1.9 THz and 40 µW when the last two triplers are cooled down to 80K. The LO outputs are combined via a wire grid inside the LO cryostat.

#### 3) LO cooler design

The two solid state chains are integrated with the control electronics power supplies and closed-cycle cooling machine in a compact module (Fig. 4-5). We use the AIM SF100 cooler, which is a Stirling cryocooler based on a moving magnet. It provides 2 W of cooling power at 80 K, consuming 75 W of electrical power. The last two triplers of each VDI LO chains are placed inside a small cryostat. The cold tip from the cooler is thermally connected to the triplers via flexible copper straps. To thermally isolate the triplers from the room temperature components, but keeping low insertion losses, 1" stainless steel WR-5 waveguide pieces are placed between the last room temperature doubler and the first cooled tripler. The measured loss of such a waveguide is about 1.5 dB at 230 GHz. To further decrease these losses, the National Radio Astronomy Observatory (NRAO), Charlottesville, VA, USA, copper plated those waveguides. After plating with more than 10 µm thick copper, the losses at 230 GHz for the 1" long waveguides decreased to ~0.5 dB.

The small LO cryostat optical window is a 1 mm-thick high resistivity Silicon window from Tydex, LLC, coated on both sides with Parylene. The window is optimized for best transmission at 1.9 THz (above 88 %).

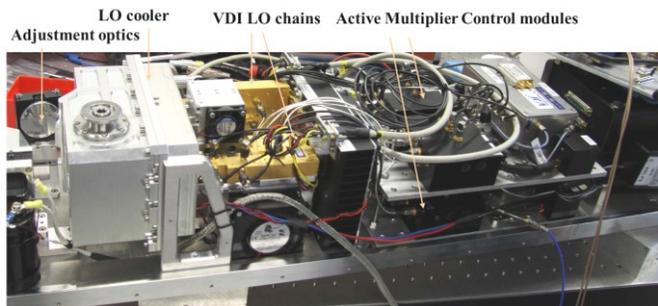

Fig. 4. LFA Local Oscillator box assembly. The LO cryostat and warm adjustable optics are shown on the left hand side. The active multiplier control modules provide the required bias to the various multipliers. The last two triplers of each chain, placed in the LO cryostat, are self-biased.

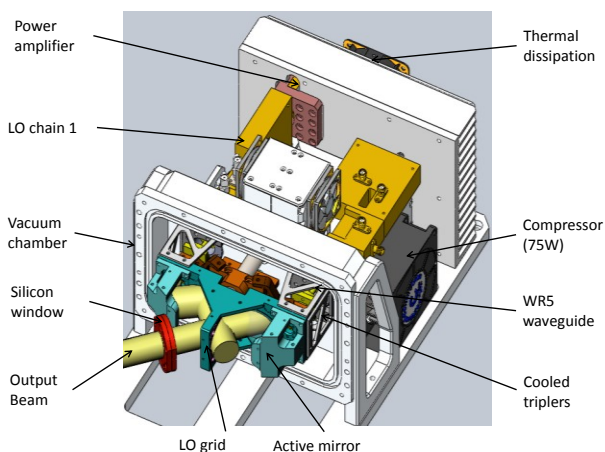

Fig. 5. View of the VDI LO chains together with the LO cooler (one half cover is removed). The combined LO output signal can be seen. The two LO chains are placed with 90 degrees polarization difference and the LO grid combines them at its output. The RF window is a 1 mm-thick Silicon with Parylene-C coating.

*4) Reference synthesizer*

The solid state multiplier chains need a +10 dBm reference signal in the 13.1-13.4 GHz range (multiplication factor is 144). The characteristics of the reference signal are critical, as the final phase and AM noises of the system strongly depends on it. Various types of high performance lab bench synthesizers were successfully used to drive VDI multiplier chains for the GREAT single pixel receivers. However, for the upGREAT development, due to space and weight limitations on the instrument mounting structure, a more compact solution was needed. Several synthesizer modules were tested (VCO-based and YIG-based) covering 8-20 GHz. When comparing the overall receiver performance, the best results are achieved with the VDI synthesizer. The receiver sensitivity compares then very favorably to the lab-bench synthesizers (equal or better performance). Using other models, such as VCO-based synthesizers, the receiver noise temperature can degrade by up to 20%. We therefore selected the VDI synthesizers to drive the GREAT/upGREAT receivers.

*C. HFA array Local Oscillator*

The local oscillator source for the HFA receiver at 4.7 THz will be a quantum cascade laser local oscillator. A prototype has already been demonstrated [30], having more than 100 µW output power, intrinsic linewidths below 10 MHz and is routinely used with the GREAT H-channel receiver. A parallel development is ongoing between the groups KOSMA at the Universität zu Köln and the DLR-Pf Berlin, for the final upGREAT HFA local oscillator.

*D. Frontend Cryostat and closed-cycle cooler*

The cryostats accommodating the upGREAT mixer arrays are partly based on the GREAT cryostats design. Both have the same mechanical interface to fit the mechanical structure which connects to the telescope flange (see Fig. 1). Therefore, the external dimensions and mass for the upGREAT cryostats are severely constrained and must stay within fixed mechanical limits. CryoVac GmbH fabricated the upGREAT cryostats using the same process specifications as for the GREAT cryostats, which had already been certified for airworthiness.

The major difference between the GREAT single pixel receivers and the upGREAT multi-pixel receivers is the cooling mechanism. The first generation of cryostats uses liquid nitrogen and liquid helium for cooling. However with 7 or 14 detectors, the cryostat space is limited and power dissipation in the active components (LNAs) becomes too large in order to keep hold times above 15 h at 4.2 K (which is the minimum required onboard SOFIA for a full night of observations). This would require having much larger liquid $N_2$ and He tanks and then space and weight limitations become prohibitive to accommodate the arrays. It was therefore decided to use closed-cycle pulse tube coolers.

*1) Pulse Tube cooler*

The receivers are cooled by closed-cycle pulse tube refrigerators, model PTD-406C from TransMIT GmbH [31]. The pulse tube was custom modified to accommodate a small 0.2 L helium pot, connected directly to the $2^{nd}$ stage. It uses a separate helium circuitry, which allows liquefying He [32]. The minimum quantity of liquid helium to suppress almost completely the temperature fluctuations of the pulse tube was found to be 30 cm$^3$. The dampening is about a factor of 25, reducing peak-to-peak modulation of 300 mK, to less than 12 mK at the cold head.

One of the main selection criteria for the cooler was its ability to withstand large tilts with minor impact on the cooling performance. This model PTD-406C has very long tubes, which are less sensitive to convection losses. The specification was to be able to tilt the cooler to ± 45° orthogonal to the telescope elevation axis with minor cooling degradation.

The pulse tube cooler provides about 0.8 W of cooling power at 4.2 K when using the Sumitomo air-cooled compressor CSA-71A which is the selected compressor to operate the LFA 14-pixel receiver. With the smaller Sumitomo air-cooled compressor (CNA-31C), the cooling power at 4.2 K is about 0.5 W, sufficient to be used with the HFA 7-pixel array.

## 2) upGREAT cryostats design

The cryostats' main components are shown in Fig. 6. There are two cooling stages, at ~40K and ~3K. The pulse tube is connected to its rotary valve separated by a 0.75 m flexible helium line. This allows decoupling the rotary valve mechanical vibrations from the detectors if desired. In our case, the rotary valve is rigidly attached to the cryostats.

The pulse tube cooling stages are thermally connected to the $1^{st}$ and $2^{nd}$ stage plates through flexible copper straps. Fiber glass structures provide the mechanical support for these stages and their components. They consist of 2 mounting flanges or rings with a G10 cylindrical structure in between them.

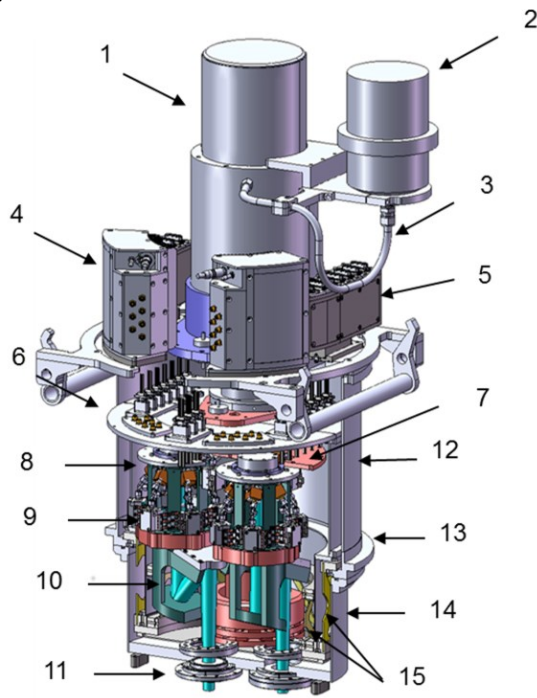

Fig. 6. upGREAT cryostat main components. 1) Pulse tube PTD-406C cold head. 2) Rotary valve. 3) Flexible Helium line. 4) Warm IF amplifiers. 5) Pre-amplifiers for biasing HEBs. 6) 40K stage plate. 7) 4K stage plate. 8) HEB detectors. 9) Cryogenic SiGe amplifiers. 10) Cold optics. 11) RF windows. 12) Vacuum Vessel upper part. 13) Seal ring. 14) Vacuum Vessel lower part. 15) Fiber glass Supports 4K-40K.

The outer vacuum walls of the upGREAT cryostats are separated into two parts, the lower one incorporating the vacuum RF windows. The windows are made of high resistivity 525 µm-thick Silicon with etched linear grooves on both sides to act as matching layers at the wavelength of interest. They have above 90 % transmission in the 1.9-2.5 THz range and 4.7 THz range [33]. For the LFA cryostat, the infrared filters used are one layer of Zitex G104, cooled to 40 K. For the HFA cryostat, the infrared filter is a QMC dichroic low pass filter with a 8 THz cutoff frequency.

For the electrical wiring inside the cryostats, Lakeshore Phosphor-Bronze AWG-32 and AWG-36 dual-twist and quad-twist wires are used. To provide filtering, several sub-D and micro-D capacitive filter networks connectors are used, placed on the 300 K and 40 K interfaces. The top plate accommodates the electrical feedthroughs (bias connectors, temperature and heater connectors and the coaxial SMA IF connectors). Pre-amplifier bias modules and warm IF amplifiers are directly mounted on this top plate (items 4-5 in Fig. 6).

## 3) Cryostats Cooling Performance

The summary of the cooling performance for the cryostats using various compressors is presented in Table IV.

TABLE IV
upGREAT COOLING PERFORMANCE

| Compressor | CNA-31C | CSA-71A |
|---|---|---|
| Compressor input power | ~3.5 kW | ~8.0 kW |
| Cooling Power $1^{st}$ stage | 10 W @60 K | 10 W @45 K |
| Cooling Power $2^{nd}$ stage | 0.5 W @4.2 K | 0.8 W @4.2 K |
| Cold end temperature | 3.0 K | 2.8 K |
| Mixer physical temperatures on blocks | 4.4 K average with 7 pixels (HFA band) | 4.2 K average with 14 pixels (LFA band) |
| Temperature Fluctuations | ±150 mK on base plate | ±150 mK on base plate |
| | ±6 mK with >30 cm$^3$ LHe in Helium pot | ±6 mK with >30 cm$^3$ LHe in Helium pot |
| Cooling time | 24 h to 3.5 K | 17 h to 3.3 K |
| | 60 h to 3.0 K | 60 h to 2.8 K |
| Tilt impact (±45°) | < 200 mK change on $1^{st}$ stage | < 200 mK change on $1^{st}$ stage |
| | < 1 mK change on $2^{nd}$ stage | < 1 mK change on end stage |

For the 14-pixel LFA cryostat, the minimum base plate temperature is 2.5 K when the LNAs are turned off, and increases to 2.8 K when they are nominally biased. The temperature on the detectors is about 4.2 K when using the CSA-71A compressor with 14 pixels, and 4.4 K when using the CNA-71A compressor with 7 pixels. The tilting impact was verified by rotating the whole assembly including the cryostat by ± 45°. This confirmed that the $2^{nd}$ stage temperature stays stable within 1 mK and the $1^{st}$ stage temperature within < 200 mK.

The temperature fluctuations measured directly on the $2^{nd}$ stage plate were confirmed to be drastically minimized when liquefying Helium in the Helium pot. However, the measured temperature fluctuations on the mixer blocks were small enough (below 1 mK peak to peak) so that using the external Helium circuitry for liquefaction was not necessary in the final flight configuration.

## E. Optics Design
### 1) Common Optics

TABLE V
SOFIA - OPTICAL PARAMETERS

| | |
|---|---|
| Primary diameter | 2500 mm |
| Primary focal length | 3200 mm |
| Telescope focal length | 4914 mm |
| Blockage | 522 mm |

The main SOFIA telescope optical characteristics are summarized in Table V. Details on the telescope common optics can be found in [21]. The SOFIA telescope nominal waist position is located 300 mm away from the SI flange interface, towards the instrument. With the upGREAT arrays, a mirror-type derotator is used to compensate for the sky rotation in the focal plane of the telescope, which happens on this type of Alt-Az mount telescopes. It consists of 3 optical quality flat mirrors mounted on a rotational stage. The derotator is computer controlled and can compensate for the sky rotation. With this, the pixel pattern can be kept fixed on sky (in the equatorial or instrument reference systems) or rotated as desired, allowing for different, sophisticated observing strategies.

*2) LFA Coupling Optics*

TABLE VI
upGREAT – DESIGNED OPTICAL PARAMETERS

|  | LFA | HFA |
| --- | --- | --- |
| RF Bandwidth | 1.9-2.5 THz | 4.7 THz ± 4 GHz |
| Edge Taper | 13 dB | 13 dB |
| Beam waist at SOFIA focal plane | 2.4 mm at 1.9 THz<br>1.7 mm at 2.5 THz | 1.0 mm |
| Half Power Beam Width (HPBW) | 15.5" at 1.9 THz<br>11.8" at 2.5 THz | 6.3" |
| Pixel spacing on sky | 34.0" | 13.8" |

The main designed optical parameters of the LFA array are summarized in Table VI. The telescope secondary edge taper was chosen to be 13 dB. The pixel spacing is then 2.2 x HPBW at 1.9 THz corresponding to 34 arcseconds on sky. The beam spacing at the mixer focal plane position is 12 mm, and the beam spacing at the SOFIA focal plane position is 8 mm, therefore the magnification factor was designed to be 1.5. The optical design consists of a pair of Gaussian telescopes, of magnifications 1.35 and 1.1. The optical design is solely based on reflective optics (no lenses are used). The main components are shown in Fig. 7.

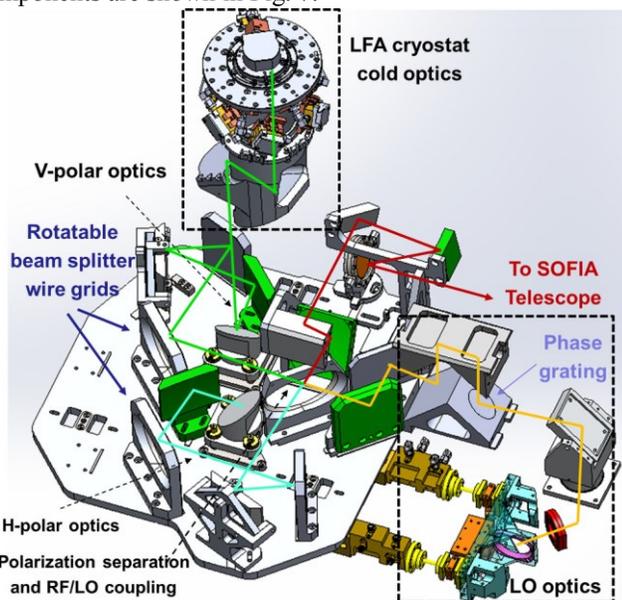

Fig. 7. View of a main optical components. The LO beam path from the 2 VDI chains is shown in orange, the input RF signal from SOFIA is shown in red. The combined RF/LO signals for both polarizations are shown in green and light blue. The active mirrors are drawn in green.

The first LFA optical design (Fig. 8) is based on diplexers to allow using local oscillators with limited output power, which is the case for frequencies above 1.9 THz, where LO sources are scarce and limited. A description of this type of optical design for array receivers using a pair of Gaussian telescope and diplexers can be found in [8, 12].

A second optical plate was manufactured identical to the first design replacing the diplexers with beam splitter wire grids and including additional mirrors to keep the same overall path length. This is of interest when there is sufficient LO power available. The wire grids can be manually rotated to vary the LO coupling from 15% to 40% while also allowing the usage of the full IF bandwidth of the HEB mixers.

The main advantage of a diplexer-based LO coupling optics is that all of the LO power is effectively used. The limitations are that it is very difficult to achieve identical path-lengths for all offset pixels, causing sensitivity degradation [34], the IF bandwidth is limited to at best 1.5 GHz (having the most sensitive lower IF blocked by the diplexer) and the added complexity of diplexers often cause receiver instabilities. Therefore, depending on the amount of LO power available, either the diplexer or beam-splitter coupling optics can be used.

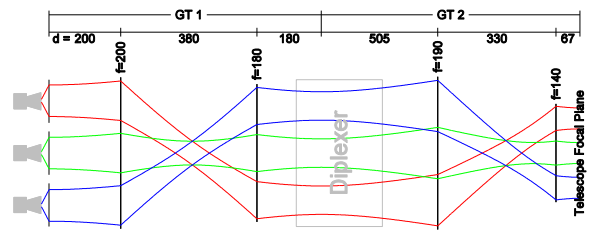

Fig. 8. Schematic beam path using two Gaussian telescopes for reimaging. The LO diplexer can be placed if needed between them. The cryostat window is located at an image of the telescope aperture, where the total beam cross section is minimal. For compactness and clarity, imaging elements are drawn as lenses, although mirrors are more commonly used in real instruments

*3) HFA Coupling Optics*

The HFA optics design is a scaled version of the LFA beam splitter coupling optics. The optical design of the HFA is summarized in Table VI (last column). The spacing between the pixels is 2.2 x HPBW at 4.7 THz (edge taper of 13 dB). The beam spacing at the mixer locations is 10.5 mm, and the beam spacing at the SOFIA focal plane position is 3.5 mm, therefore the magnification factor is designed to be 3.0. The optical design consists of a pair of Gaussian telescopes, each of magnification 1.5 and 2.0.

*F. HEB Mixers*

The superconducting detectors are Hot Electron Bolometers developed by the KOSMA group at the Universität zu Köln, with state of the art performance. The NbN HEB mixers fabrication used for upGREAT is described in [23, 35].

*1) LFA HEB mixers*

The HEB mixers designed for the LFA receiver cover the complete RF bandwidth from 1.9-2.5 THz using a broadband single-sided waveguide probe antenna and CPW transmission lines matching circuits. The waveguide and device recess features of the metal block are fabricated by a procedure involving multiple stamping and milling steps and resulting to micron-precise feature reproducibility. The devices are fabricated with 3.5 nm NbN films on 2 μm thick Silicon substrate and beamleads are used for contacts. Several wafers were fabricated having LFA devices of various types.

Fig. 9 shows a scanning electron microscope (SEM) picture of one of such devices. For upGREAT, deposition of the NbN

layer was further optimized to thinner layers in order to meet the LO requirements of the array. In comparison to the devices fabricated for the GREAT H channel [23], the LO power requirement was reduced by a factor of 3. HEB microbridge dimensions where reduced from 5.5 x 300 x 3600 to 3.5 nm x 200 nm x 3250 nm with an impedance required by the on-chip circuit of 120 Ohms. The typical $T_c$ of the HEBs are 8.1 ± 0.3 K. The horns are electroformed and are spline smooth-walled horns, designed and fabricated by Radiometer Physics GmbH [36].

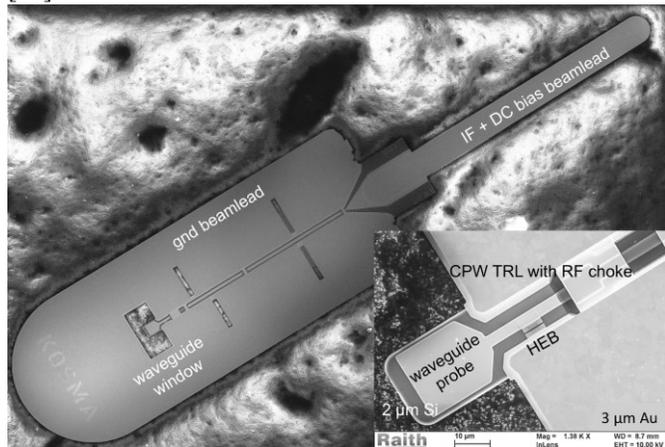

Fig. 9. View of a HEB NbN device for the LFA array. The ground contacts are done via the large beamleads. The IF output beamlead is bonded to a transmission line on a separate circuit board and for strain relief this TRL is wire bonded to the center conductor of a SMA connector.

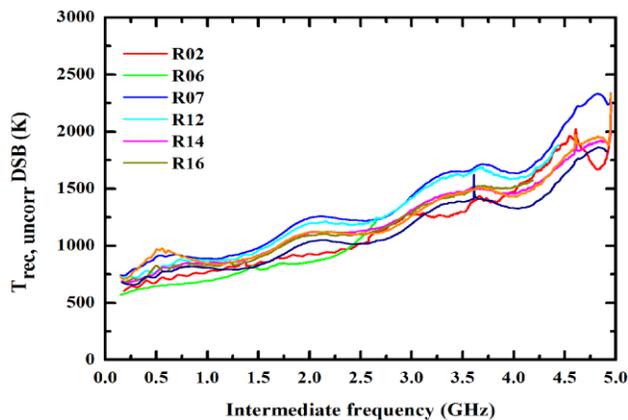

Fig. 10. Noise temperature characterization of the LFA mixers in a test cryostat. No corrections were applied with approx. 90% signal coupling and non-evacuated signal path. The devices are from one section "R" of the flight devices wafer and the mixers are used in the H polarization array.

The average performance of the HEB measured during RF qualification at KOSMA was an uncorrected DSB noise temperature of 800 K ± 50 K for the 15 mixers qualified for upGREAT and their 3 dB noise bandwidth was 3.7 ± 0.1 GHz (Fig. 10). Compared to the GREAT L2 mixer, the $T_{REC}$ are equal or better and due to the NbN layers, the IF noise bandwidth is much improved (2.3 GHz for NbTiN devices, see Fig. 16). A strong influence of on-wafer position of the devices on LO power requirement was found and is thought to be a result of the inhomogenities of the thinner NbN film. Hence for each the subarrays devices from one section were selected in order to ensure uniformity of LO power required. The so-called "R" devices were used for the H pol array and "J" devices for the V pol array. The "J" devices require 2-3 times more LO power than the R devices, which can be attributed to their higher $I_C$ values, 250 μA vs. 180 μA, respectively.

*2) HFA HEB mixers*

The mixers for the HFA receiver are based on the H-channel HEB mixer [23]. As for the LFA, the HFA devices will use the same wafer with thinner NbN as compared to the H channel mixers. For the HFA array LO power is not as sparse as with the LFA due to the QCL based LO, which provides more output power. Hence the devices selected should easily meet the requirements and allow efficient beam splitter coupling. The expected noise temperature should be comparable to the measured H-channel mixer results, which achieved 800 K DSB and an IF noise bandwidth of 3.5 GHz, comparable to the LFA mixers. The new mixers were fabricated in parallel with the LFA devices on the same wafers and the selection of the best devices and their characterization will take place end of 2015. Integration in the HFA receiver is scheduled for the 1st quarter of 2016.

*G. IF Signal Processing*

The HEB output IF signal in the range 0.5-4 GHz is amplified by a set of SiGe cryogenic low noise amplifiers CITLF4 [37], which have nominally a low thermal dissipation of 12 mW. This can be lowered, if required, to 6 mW with little impact on the cryogenic LNA noise temperature performance. Fig. 11 shows 15 cryogenic amplifier characteristics, having a gain of 35-40 dB and a noise temperature of 4-5 K when measured at 23 K physical temperature. These LNAs also include the bias-T for the HEB mixer biasing. The default 5 K resistors provided for that purpose for the CITLF4 models were removed and replaced by custom-made coils. This improved the noise temperature by ~1 K. The cryostat output IF signals are then further amplified by low noise room temperature amplifiers from MITEQ (AFS4), and afterwards processed by room temperature components (amplifiers, filters, equalizers, variable attenuators) located in in a dedicated IF processor (Fig. 12).

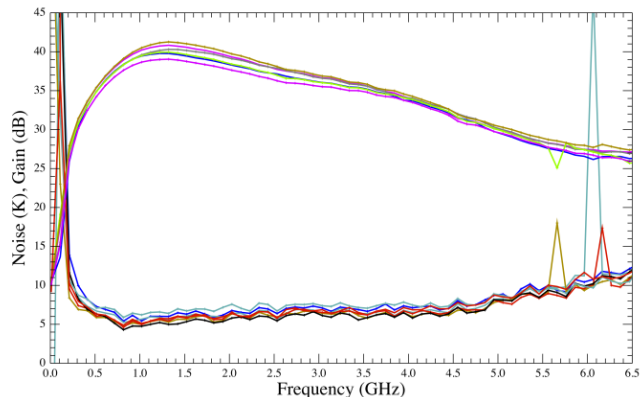

Fig. 11. Measurements for 15 cryogenic low noise amplifiers CITLF4 after modification and inclusion of a coil-based bias-T for biasing of the HEB mixers. The gain when measured at 23 K is between 35-40 dB and the noise temperature is 4-5 K between 0.5-4 GHz of IF bandwidth. The resonances above 5 GHz are due to the coil.

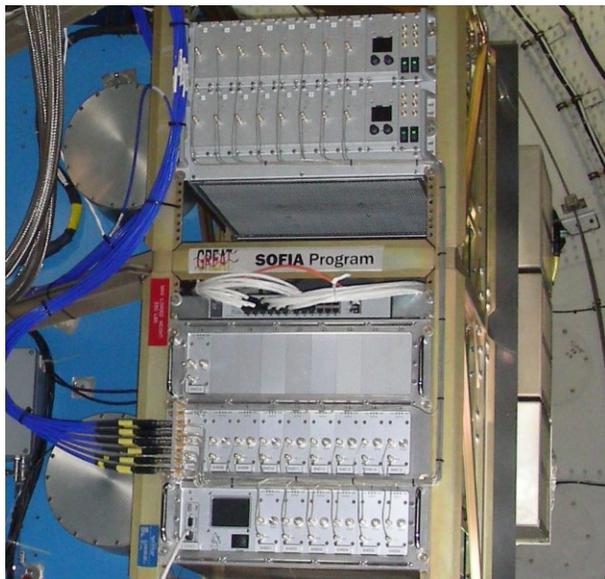

Fig. 12. The upGREAT IF processor, located in the lower half, contains up to 22 cartridges. Each of them is amplifying, filtering, equalizing and performing remote control leveling of its IF power to provide the nominal input power to the spectrometers. The upper half of the picture shows the FFTS-4G spectrometers. In this case 16 boards are installed.

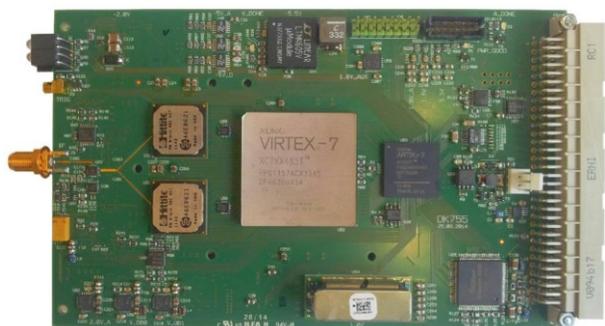

Fig. 13: The FFTS-4G boards, using the VIRTEX-7 FPGA, two Hittite ADCs with 6 GHz input bandwidth are interleaved. Each board is capable of instantaneously sampling the 0-4 GHz band with up to 64K spectral channels. For the upGREAT receivers, the number of channels is 32K.

Finally, the signals are fed into an MPIfR-built Fast Fourier Transform Spectrometers with 4 GHz of instantaneous bandwidth (FFTS-4G, Fig. 12-13). They are conceptually based on the 0-2.5 GHz FFTS spectrometers [38-39] and this newer generation is able of instantaneously sampling the 0-4 GHz range using its base band (but can also sample the 4-8 GHz region if using its 1st Nyquist band).

Each FFT-4G board uses two Hittite 8-bit ADCs (HMC5448) which are time interleaved. The number of channels used for upGREAT is 32K, obtaining a channel spacing of 122 kHz with an equivalent noise bandwidth (ENBW) of 142 kHz.

## IV. LFA RECEIVER INTEGRATION AND LABORATORY RESULTS

### A. Focal plane arrays and cryostat components

The final integration of all LFA HEB mixers in the cryostat took place in February 2015. A mixer sub-array is shown in Fig. 14, with the 6 offset mixers in place. The inner cryostat assembly can be seen in Fig. 15. A hexagonal grid is chosen for the upGREAT arrays to maximize the packing density and thus the mapping efficiency for compact sources.

The offset mixers are located around aluminum towers providing support for the mirrors and cryogenic low noise amplifiers. The IF outputs are then connected via flexible coaxial lines to feedthrough SMA connectors, and then connected via stainless steel coaxial conductors to the cryostat output SMA connectors.

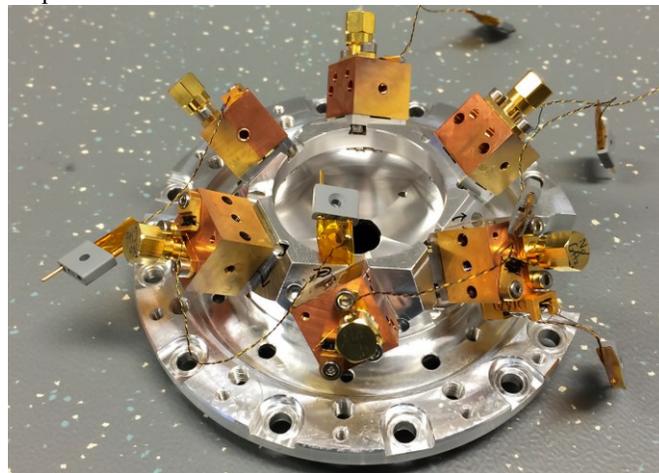

Fig. 14. One of the LFA sub-arrays HEB mixers mounted on its support structure. The central part accommodates offset parabolic mirrors, placed in a 12 mm side hexagonal configuration, defining the pixel spacing. Only the 6 offsets pixels are mounted there and the central pixel has a separate mounting structure not shown in the picture.

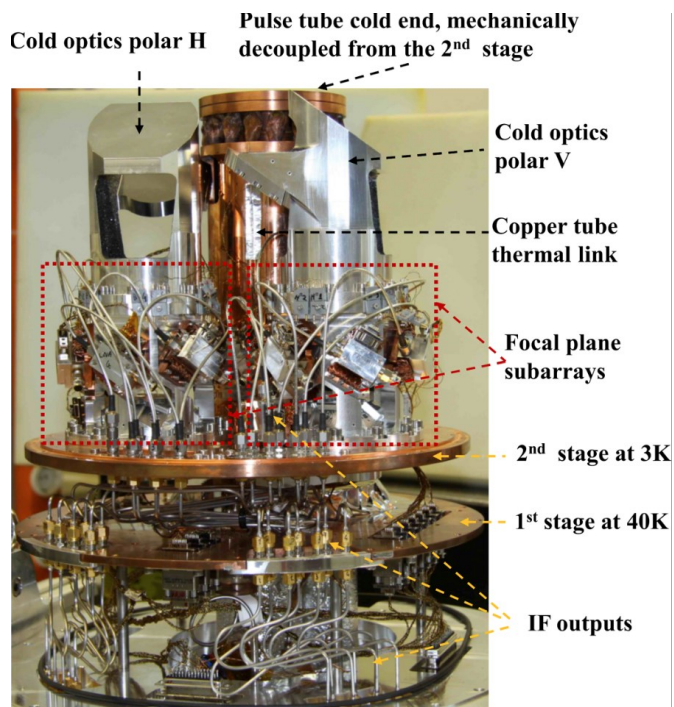

Fig. 15. LFA cryostat 4K components. The two sub-arrays can be seen, located symmetrically with respect to the copper heat sink tube, which provides cooling to the 4K stage plate. The mixers are located inside the aluminum towers and surrounded by the cryogenic low noise amplifiers. The IF outputs are then connected via flexible coaxial lines to feedthrough SMA connectors, and then connected via stainless steel coaxial conductors to the cryostat output SMA connectors.

### B. Receiver characterization

#### 1) LO coupling

As the LO multiplier chains for the LFA receiver provided sufficient power at the [CII] frequency (1.905 THz), the preferred coupling optics was using beam-splitter wire grids rather than diplexers. Both optics were tested but the best results were achieved with the beam-splitter coupling scheme. However, strong interferences between the two LO chains when operated in parallel, impeded the simultaneous use of both sub-arrays during the first commissioning.

#### 2) Noise Temperature Measurements

When using the beam splitter optics plate, the H-polarization sub-array which carries the R-section HEB devices needed to have about 18% of the LO power coupled. The other polarization (V) using the J-sector devices needed about 40% LO coupling.

We measured the noise temperature of the whole receiver, evacuating on the optical compartment to simulate the flight conditions at an altitude above 40000 feet. We use an absorber cooled to liquid nitrogen temperature as a cold load and a SiC with Stycast hot load absorber (loads at 77 K and 295 K) [22].

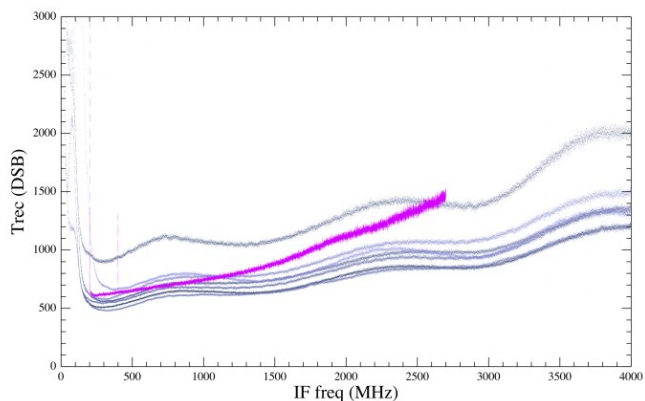

Fig. 16. Uncorrected $T_{REC}$ DSB (K) at 1.9 THz versus IF frequency for the H-polarization LFA 7 channels. The noise IF bandwidth is about 3.3 GHz. The minimum noise temperature (DSB) for the best 6 pixels is between 500-700 K. One mixer is underpumped and therefore that channel noise temperature degrades to about 1000 K. As a comparison, the performance of the single pixel L2 receiver is shown in purple. Its noise IF bandwidth is smaller (~2.5 GHz) as it uses a NbTiN HEB mixer.

The uncorrected noise temperatures include all the contributions from the IF chain, HEB mixer, infrared filter, RF window, and signal/coupling losses (shown in Fig. 16 and summarized in Table VII). The noise IF bandwidth is about 3.5 GHz. The minimum noise temperature (DSB) for the best 6 pixels of the H-polarization array is between 500 - 700 K. One mixer is underpumped and therefore that channel noise temperature degrades to about 1000 K. As a comparison, the performance of the single pixel GREAT L2 receiver is shown in purple in Fig. 16.

For the V-polarization array, the uncorrected noise temperatures at 1.9 THz are about a factor of 2 worse than the H-polarization, with minimum noise temperatures (DSB) of about 1300 K. This is due to the higher LO coupling required to pump the mixers (40% coupling needed). We are currently replacing the 2nd polarization mixers with devices requiring less LO power which will result in an improved signal coupling.

TABLE VII
upGREAT – LFA SENSITIVITIES

|  | H-Polarization | V-Polarization |
|---|---|---|
| LO Bandwidth | 1.88 – 1.92 THz | 1.88 – 1.92 THz |
| $T_{REC}$ (DSB) | ~600K at 500 MHz IF [1] ~1200K at 3500 MHz IF | ~1300K at 500 MHz IF [2] ~2600K at 3500 MHz IF |
| LO coupling | Beamsplitter– 18% coupling | Beamsplitter– 40% coupling |

[1-2]Performance for 6 out of 7 pixels, the 7th pixel is underpumped and is about 40% noisier

#### 3) Beam Characterization

Beam characterization and alignment are performed in an iterative fashion. In the MPIfR laboratory in Bonn, optical and radio alignment were made by using a dedicated rotating wheel which allows deriving the average positions for all the pixels in a location close to the telescope focal plane. Then far distance beam measurements at distances ranging from 500 mm to 2000 mm from the telescope waist position were performed. Beam propagation direction and beam characteristics can then be derived for the individual pixels. By slightly adjusting the cold optics and verifying the applied changes, the beam characteristics were confirmed to be sufficiently close to the designed values.

#### 4) Receiver Stability

Another important figure of merit for the receivers is the total power and spectroscopic stabilities. This will directly impact on the quality of the observations (e.g. unstable bandpass renders line profiles analysis more difficult).

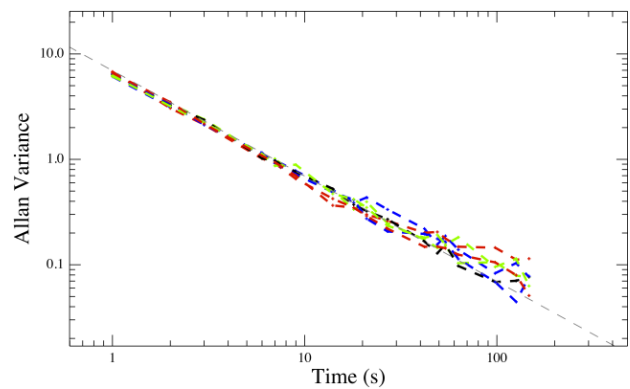

Fig. 17. Example of spectroscopic Allan Variance performed at an LO of 1.902 THz. The compared spectrometer channels have a 1.4 MHz effective bandwidth and are spaced by 700 MHz. Typical Allan times are above 80s.

There are numerous potential factors than can impact and limit the receiver stability, for example: pulse tube cooler temperature fluctuations and vibrations, telescope elevation changes, microphonics in the optical compartment, LO cooler temperature fluctuations and vibrations, LO chains output power stability, VDI driver synthesizer output power stability, HEB gain stability, IF components gain variations, spectrometer stability, etc. A convenient way to assess the overall total power and spectroscopic stability of a receiver is using the Allan variance [40]. We characterized the LFA receiver at various frequencies and bias conditions and the measurements show low total power Allan variance times (3-5s) in a 1.4 MHz measurement bandwidth. The spectroscopic Allan variance times comparing 1.4 MHz spectrometer channels spaced by 700 MHz are better than 80-100 seconds (Fig. 17).

The stability is probably slightly worse than the single pixel GREAT receivers which had spectroscopic Allan times above 100 seconds, after optimization over several observing campaigns. The current performance is already sufficient for efficient observations. By knowing the receiver stability values, the observations modes can be optimized to provide best quality baselines.

## V. INSTALLATION AND COMMISSIONING ONBOARD THE SOFIA OBSERVATORY

### A. LFA receiver installation

After the successful LFA receiver characterization in the MPIfR facilities, the receiver and all of the associated components and electronics were sent to the NASA Armstrong Flight Research Center (AFRC), Palmdale, CA, USA in March 2015, where the SOFIA airplane is based. The final characterization was performed in those laboratory facilities using the GREAT mechanical structure, which is the part interfacing with the telescope (Nasmyth tube attachment).

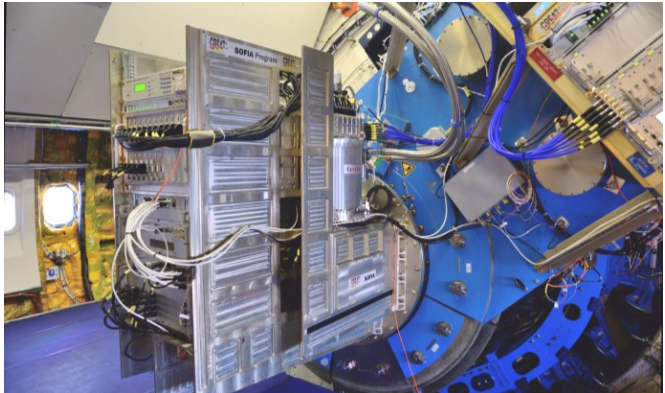

Fig. 18. The GREAT/upGREAT instrument mounted on the Nasmyth tube of the SOFIA telescope. On the left side, the electronics rack houses the sensitive bias electronics, reference synthesizers and other motor controllers. The LFA cryostat can be partly seen on the right hand side of the central instrument structure. The pressurized helium hoses can be seen going from the rotary valve on the cryostat to the top part. 15 coaxial blue lines (14 for the LFA receiver and 1 for the L1 receiver) connect the cryostats IF outputs to the IF processors and spectrometers FFTS-4G.

The installation of the upGREAT system aboard the SOFIA observatory took place on May 1$^{st}$ 2015 (Fig. 18-19). The chosen configuration for GREAT/upGREAT was to use in parallel the single pixel L1 receiver (1.25-1.5 THz) together with the LFA receiver. The telescope signal is separated by a QMC low-pass filter dichroic (W1449), letting the 1.25-1.5 THz pass in transmission (95%) and reflecting the frequencies above 1.8 THz (>99%)

After the initial installation onboard SOFIA, the instrument functionalities were verified. Beam measurements close to the telescope focal plane were repeated and an alignment check was performed at the sub-reflector, using $LN_2$-cooled absorber material to ensure that the receivers L1-LFA were co-aligned and well centered.

For the commissioning flights, both LFA sub-arrays could not be operated in parallel as the LO chains were causing strong interferences. The H-polarization sub-array was therefore selected, having the best performance. The V-polarization sub-array was used partially during one flight to verify its on-sky performance and co-alignment with the H-array.

### B. Cryocooler Infrastructure aboard the SOFIA airplane

In parallel to the development and testing of the upGREAT instrument in the laboratory facilities, the SOFIA project designed, built, integrated and tested the required associated infrastructure to operate the Pulse Tube cryo-coolers aboard the aircraft in 2014-2015, with inputs from MPIfR and co-funded by DLR.

Perhaps the most demanding aspect of this work package was the certification of the various components for airworthiness. Such cryocooler systems are very common on ground-based telescope, or hospital magnetic resonance imaging facilities, but are not intended to be operated onboard an airplane. The strong accelerations, turbulence, and tilt angles can cause oil displacement and contamination that compromise the cooling performance of the pulse tube cold head, or might damage the compressor capsule itself in extreme cases. Following recommendations from Sumitomo, it was decided to mount the compressor onto a vibration-isolated, single axis Gimbal mount, to compensate for the airplane accelerations/decelerations in the flight direction and pitch axis tilt. Also, a molecular sieve filter and an additional oil adsorber are mounted externally to the compressor to further reduce the chances of oil contamination reaching the Pulse tube during strong turbulent events.

After developing Concept of Operations and System Specification documents, the NASA Ames group built, assembled and certified the various components comprising this system through a thorough design, testing and validation program. From the pressurized Helium lines, adding up to 35m length for both return and supply side, to the custom made manifold, which includes various pressure relief valves and burst discs, and the compressor itself, all components and assemblies were certified for compliance with the applicable NASA standards for structural integrity, pressure systems safety, environmental acceptance and airworthiness.

The compressor and control electronics enclosure underwent comprehensive development and environmental acceptance testing (vibration, temperature, air pressure). Initial low amplitude sine sweep vibration testing of the compressor revealed several components and assemblies with natural frequencies that could be excited by aircraft vibrations and lead to potentially damaging resonance, given the damping characteristics of the vibration isolation mounts. These components and assemblies were stiffened to ensure that all natural frequencies were greater than 20 Hz to ruggedize the compressor. The electronic control system incorporates various warning and alarm signals, for the following parameters: roll and pitch angles, vertical and lateral accelerations, various temperatures monitored inside and outside the compressor, and the supply and return pressures of the Helium lines. Exceeding the warning thresholds triggers annunciators to notify the operators, and if any of the alarm values is reached, the compressor automatically switches off.

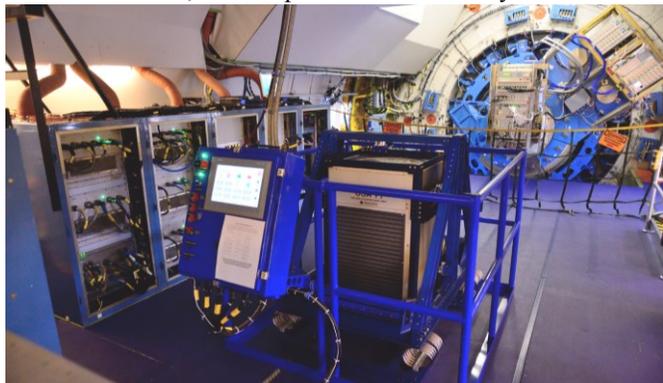

Fig. 19. View from inside the SOFIA aircraft, looking in the aft direction. In the foreground, the Sumitomo CSA-71A compressor can be seen in a vibration-isolated, one-axis Gimbal mount. The electronic control enclosure with a touchscreen panel is mounted to the guard rail. In the background, the GREAT/upGREAT instrument is mounted to the SOFIA telescope's Nasmyth tube flange.

*C. Commissioning flights results*

From May 12$^{th}$ to 22$^{nd}$ 2015, a series of four commissioning fights was successfully performed. The LFA receiver performed smooth observations during these flights. The receiver sensitivities were as measured in the laboratory, the cryostat 2$^{nd}$ stage temperatures were stable to within 1 mK, independent of the telescope elevation and flight conditions. The commissioning and science observations were of excellent quality, confirming the state of the art sensitivities. Various observing scenarios were performed to validate the inflight performance of the array. An example of a large scale on-the-fly mapping of [CII] in S106 is shown in Fig. 20.

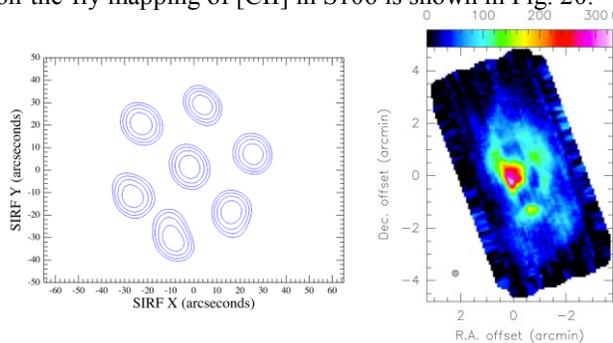

Fig. 20. *Left*: Preliminary beam maps taken on Saturn, showing the positions of the beams on sky for the H-polarization array. Contours are spaced every 10%. *Right*: One of the first observations of the S106 nebula using the LFA receiver on May 21$^{st}$, measuring the emission line of ionized atomic carbon [CII] at a wavelength of 158 µm. The effective angular resolution is shown by the circle in the lower left corner.

The LFA receiver main commissioning results (derived characteristics such as beam efficiencies, beam sizes, observing modes summary, frequency and intensity calibrations) will be described in [41].

These first series of flight using the LFA receiver also provided substantial data about the associated cryocooler infrastructure. This was essentially a confirmation that the designed system performed nominally. The Gimbal mount allowed operating the compressor during the take-offs, the climbing phase and the entirety of the flights, except for the landing phases, where the compressor was preventively switched off.

## VI. CONCLUSIONS

We presented the design of the upGREAT THz heterodyne arrays for frequencies of 1.9-2.5 THz (2x7 pixels in orthogonal polarizations) and 4.7 THz (7 pixels in a single polarization). The first array was successful built, tested and commissioned in May 2015 at the [CII] frequency (1.905 THz) with the SOFIA airborne observatory. The performance is state of the art, with ~600 K minimum uncorrected DSB $T_{REC}$, with a 3 dB noise IF bandwidth of 3.5 GHz for one polarization. The 2$^{nd}$ polarization has ~1300 K minimum uncorrected DSB $T_{REC}$. We are currently replacing the 2$^{nd}$ polarization mixers with devices requiring less LO power and its performance should then become as good as for the 1$^{st}$ polarization. Also, a modified LO optical scheme is being implemented, which will allow operating the 14 pixels in parallel. This will be available for the next flight series in December 2015.

A following step will be to employ different local oscillator sources to cover the full 1.9-2.5 THz frequency range. The HFA array receiver at 4.7 THz is currently being integrated and is expected to be commissioned in 2016.

## NOTE ADDED IN PROOF

In December 2015, a new flight series with SOFIA has taken place where the 2$^{nd}$ polarization mixers were replaced achieving improved sensitivities (similar to the 1$^{st}$ polarization mixers) and the optical scheme was modified, allowing us to successfully operate the 14 mixers in parallel.

## ACKNOWLEDGMENT

The authors thank Sener Türk from the microwave group in MPIfR, Bonn, Germany, for fabricating the stainless steel WR-5 waveguide assemblies used in the local oscillator cooler. We are also very grateful to Tony Kerr and Gerry Petencin from NRAO, Charlottesville, VA, USA, for the copper plating of those waveguides. We wish to thank also the strong and constant support of Virginia Diodes Inc. Charlottesville, VA, USA. The KOSMA authors would like to thank Michael Schultz for his mechanical design work on the mixer blocks and his perseverance and precision in mounting a large number of HEB devices. Finally, the authors would like to thank the SOFIA Program Office and the staff of the NASA Ames Research Center (ARC) in Moffett Field, CA, and the Armstrong Flight Research Center (AFRC) in Palmdale, CA, for their strong continuous support during the upGREAT commissioning campaign.

## REFERENCES

[1] G. L. Pilbratt et al., "*Herschel Space Observatory* - An ESA facility for far-infrared and submillimetre astronomy", *Astron. And Astrophys.*, Vol. 518, L1, July 2010.

[2] S. W. Holland et al., "SCUBA-2: the 10 000 pixel bolometer camera on the James Clerk Maxwell Telescope", *Mon. Not. R. Astron. Soc.*, 430, April 2013.
[3] G. Siringo et al., "The Large APEX BOlometer CAmera LABOCA", *Astron. And Astrophys.*, Vol. 497, 945, 2009.
[4] C. Ferkinhoff et al., "ZEUS-2: a second generation submillimeter grating spectrometer for exploring distant galaxies", *Proc. of the SPIE* 7741, 77410Y, 2010.
[5] W. Zhang et al., "Quantum noise in a terahertz hot electron bolometer mixer", *Appl. Phys. Lett.* 96, 111113 (2010).
[6] J. L. Kloosterman et al., "Hot electron bolometer heterodyne receiver with a 4.7-THz quantum cascade laser as a local oscillator", *Appl. Phys. Lett.* 102, 011123 (2013).
[7] C. Groppi and J. Kawamura, "Coherent Detector Arrays for Terahertz Astrophysics Applications", *IEEE Trans. on THz Sci. and Technol.*, Vol. 1, Issue: 1, pp 85–96, Sept. 2011.
[8] U. U. Graf, C. Honingh, K. Jacobs, J. Stutzki, "Terahertz Heterodyne Array Receivers for Astronomy", *J. Infrared Millim. THz Waves*, Vol. 36, Issue 10, pp.896-921, 2015.
[9] H. Smith et al., "HARP: A submillimetre heterodyne array receiver operating on the James Clerk Maxwell Telescope", *Proc. SPIE*, Vol. 7020, no. 70200Z, pp. 1–15, 2008.
[10] C. E. Groppi et al, "Testing and integration of Supercam, a 64-pixel array receiver for the 350 GHz atmospheric, window," *Proc. SPIE*, Vol. 7741, pp. 1–11, 2010.
[11] U. U. Graf et al "SMART: The KOSMA sub-millimeter array receiver for two frequencies", *Proc. 13th Int. Symp. Space Terahertz Technol.*, Cambridge, MA, Mar. 26–28, 2002.
[12] C. Kasemann et al., "CHAMP+: A Powerful Array Receiver for APEX", *Proc. of the SPIE*, Vol. 6275, June 2006.
[13] C. Walker et al., "The Stratospheric THz Observatory (STO)", *Proc. of the SPIE*, Vol. 7733, 22 July 2010.
[14] D. Meledin et al., "A 1.3-THz Balanced Waveguide HEB Mixer for the APEX Telescope", *IEEE Trans. Microw. Theory Techn.*, 57(1), 89, 2009.
[15] T., de Graauw et al, "The Herschel-Heterodyne Instrument for the Far-Infrared (HIFI)", *Astron. And Astrophys*, Vol. 518(2), pp 1-7, 2010.
[16] S.Cherednichenko et al., "2.5THz multipixel heterodyne receiver based on NbN HEB mixers", *Proc. of the SPIE*, Vol. 6275, June 2006.
[17] J. Kloosterman, et al. "4x1 Pixel Heterodyne Array Development at 1.9 THz", *Proc. 26th Int. Symp. Space Terahertz Technol.*, Cambridge, MA, Mar. 16–18, 2015.
[18] X.X. Liu et al, " A 2 × 2 Array Receiver at 1.4 THz based on HEB mixers and a Fourier Phase Grating Local Oscillator", *Proc. 26th Int. Symp. Space Terahertz Technol.*, Cambridge, MA, Mar. 16–18, 2015
[19] B. D. Jackson et al, "The SPICA-SAFARI Detector System: TES Detector Arrays with Frequency Division Multiplexed SQUID Readout", *IEEE Trans. on THz Sci. and Technol*, Vol. 2, pp 12-21, 2012.
[20] A. Smirnov, "Millimetron: the next step of FIR Astronomy", *Proc. 26th Int. Symp. Space Terahertz Technol.*, Cambridge, MA, Mar. 16–18, 2015.
[21] E. T. Young et al., "Early Science with SOFIA, the Stratospheric Observatory for Infrared Astronomy", *Astrophys. J. Lett.*, Vol. 749, n. 2, April 20, 2012.
[22] S. Heyminck, U. U. Graf, R. Güsten, J. Stutzki, H. W. Hübers and P. Hartogh, "GREAT: the SOFIA high-frequency heterodyne instrument", *Astron. and Astrophys.*, Vol. 542, L1, June 2012.
[23] D. Büchel et al, "4.7 THz Superconducting Hot Electron Bolometer Waveguide Mixer", *IEEE Trans. on THz Sci. and Technol.*, Vol. 5, no. 2, pp 207-214, March. 2015.
[24] R. T. Boreiko and A. L. Betz, "Heterodyne Spectroscopy of the 63μm OI line in M42", *The Astrophys. J.*, 464, L83-L86, 1996.
[25] J. R. Gao et al., "Terahertz heterodyne receiver based on a quantum cascade laser and a superconducting bolometer", *Appl. Phys. Lett.* 86, Issue 244104, 2005.
[26] H.-W. Hübers et al., "Terahertz quantum cascade laser as local oscillator in a heterodyne receiver", *Optical Express*, Vol. 13, Issue 5890, 2005.
[27] I. C. Mayorga, A. Schmitz, T. Klein, C. Leinz, and R. Güsten, "First In-Field Application of a Full Photonic Local Oscillator to Terahertz Astronomy", *IEEE Trans. on THz Sci. and Technol.*, Vol 2, Issue 4,pp 393-399, July 2012.
[28] T. Crowe, J. Hesler, S. Retzloff, C. Pouzou, G.Schoenthal, "Solid State LO Sources for Greater than 2THz", *Proc. 22nd Int. Symp. Space Terahertz Technol.*, Tucson Arizona, April 26th-28th, 2011.
[29] U. U. Graf and S. Heyminck, "Fourier gratings as submillimeter beam splitters", *IEEE Trans. Antennas and P.*, Vol. 49, Issue 4, pp. 542-546, 2001.
[30] H. Richter, M. Wienold, L. Schrottke, K. Biermann, H. T. Grahn, and H.-W. Hübers, "4.7-THz Local Oscillator for GREAT", *IEEE Trans. on THz Sci. and Technol.*, Vol. 5, Issue 4, pp 539–545, July 2015
[31] C. Wang, G. Thummes and C. Heiden, "A Two-Stage Pulse Tube Cooler Operating below 4 K", *Cryogenics* 37, pp. 159-167, 1997.
[32] G. Thummes, C. Wang and C. Heiden, "Small Scale 4He Liquefaction using a Two-Stage Pulse Tube Cooler", *Cryogenics* 38, pp. 337-342, 1998.
[33] A. Wagner-Gentner, U. U. Graf, D. Rabanus and K. Jacobs, "Low loss THz window", *Infrared Physics & Technology*, Issue 48, pp. 249–253, 2006.
[34] M. Kotiranta, C. Leinz, T. Klein, V. Krozer, H.-J. Wunsch, "Characterization of Imperfections in a Martin-Pupplett Interferometer using Ray-Tracing", *J. Infrared Millim. THz Waves*, Vol. 33, Issue 11, pp 1138-1148, November 2012.
[35] P. Pütz et al., "1.9 THz Waveguide HEB Mixers for the upGREAT Low Frequency Array", *Proc. 26th Int. Symp. Space Terahertz Technol.*, Cambridge, MA, Mar 16–18, 2015.
[36] B. Thomas et al., "1.9-2.5 THz and 4.7 THz electroformed smooth-wall spline feedhorns for the HEB mixers of the upGREAT Instrument onboard SOFIA aircraft", *Proc. 25th Int. Symp. Space Terahertz Technol.*, Moscow, April 27–30, 2014.
[37] S. Weinreb, "Design of Cryogenic SiGe Low-Noise Amplifiers", *IEEE Trans. Microw. Theory Techn.*, Vol. 55, Issue 11, 2007.
[38] B. Klein, S. D. Philipp, R. Güsten, I. Krämer, D. Samtleben, "A new generation of spectrometers for radio astronomy: Fast Fourier Transform Spectrometer", *Proc. of the SPIE* 2006, Vol. 6275, pp. 62751, 2006.
[39] B. Klein, S. Hochgürtel, I. Krämer, A. Bell, R. Güsten, "High-resolution wide-band Fast-Fourier Transform spectrometers", *Astron. and Astrophys*, Vol. 542, June 2012, L3
[40] V. Ossenkopf, "The stability of spectroscopic instruments: a unified Allan variance computation scheme", *Astron. And Astrophys., Vol.* 479, 915-926, 2008.
[41] C. Risacher et al., "The upGREAT 1.9 THz multi-pixel high resolution spectrometer for the SOFIA Observatory", *in preparation*.

**Christophe Risacher** received the Engineering M.Sc. degree from l'École Supérieure d'Électricité (Supélec), Gif-sur-Yvette, France in 1998, and the Ph.D. degree from Chalmers University of Technology, Göteborg, Sweden, in 2005. Since 1998, he has worked at various radio astronomy observatories developing novel instrumentation and supporting observations. Among those are the IRAM 30m telescope in Granada, Spain, the Chalmers University of Technology with the Onsala Observatory, Sweden, the Apex Telescope with the European Southern Observatory, the HIFI instrument with the Herschel Observatory. Since 2011, He is working at the Max Planck Institut für Radioastronomie in Bonn, Germany, and is the project manager responsible for the development of the upGREAT array receivers for the SOFIA NASA/DLR airborne observatory.

**Rolf Güsten** photograph and biography not available at time of publication

**Jürgen Stutzki** photograph and biography not available at time of publication

**Heinz-Wilhelm Hübers** photograph and biography not available at time of publication

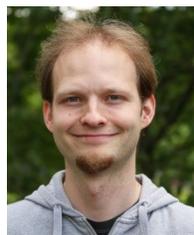

**Denis Büchel** received the B.S. and M.S. degree in physics from the Universität zu Köln, Germany, in 2009 and 2012. He is currently working since April 2012 toward the Ph.D. degree in physics at the Kölner Observatorium für Submm Astronomie (KOSMA), Universität zu Köln, Germany.

His main research interest are development and characterization of superconducting hot electron bolometer mixers for the upGREAT focal plane array extension on SOFIA.

**Urs U. Graf** photograph and biography not available at time of publication

**Stefan Heyminck** photograph and biography not available at time of publication

**Cornelia Honingh** photograph and biography not available at time of publication

**Karl Jacobs** photograph and biography not available at time of publication

**Bernd Klein** photograph and biography not available at time of publication

**Thomas Klein** photograph and biography not available at time of publication

**Christian Leinz** photograph and biography not available at time of publication

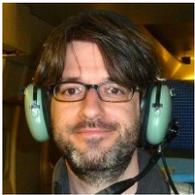

**Patrick Pütz** received the Diploma and Ph.D. degree in physics from the University of Cologne, Köln, Germany, in 1997 and 2003, respectively. He is a Senior Scientist with the Kölner Observatorium für Submm Astronomie (KOSMA), Universität zu Köln, Germany. He joined the KOSMA instrumentation group for his diploma thesis developing microfabrication processes for submicron area superconductor-insulator-superconductor (SIS) tunnel junction mixers and continued this work for his Ph.D. After graduation, he was part of the Cologne team for the Band 2 SIS mixers for HIFI on Herschel. In 2005 he joined the SORAL instrumentation group at University of Arizona, where he worked on several submillimeter and THz heterodyne receiver projects and related fabrication technology. In 2007 he re-joined KOSMA for work on 1.4 to 2.5 THz waveguide hot electron bolometer (HEB) mixers for the GREAT heterodyne instrument on SOFIA and, in 2011, the 1.9 THz mixers for the Stratospheric THz Observatory (STO). He currently is working on the HEB mixers for the upGREAT focal plane array extension for operating frequencies up to 4.7 THz. His current interests include THz waveguide and planar circuit technology for superconducting devices.

**Nicolas Reyes** photograph and biography not available at time of publication

**Oliver Ricken** photograph and biography not available at time of publication

**Hans Joachim Wunsch** photograph and biography not available at time of publication

**Paul Fusco** photograph and biography not available at time of publication

**Stefan Rosner** photograph and biography not available at time of publication